\documentclass{PoS}
\usepackage{amsmath, amssymb, graphicx,enumitem}

\newcommand{\gr}{$\gamma$-ray}
\newcommand{\grs}{$\gamma$-rays}

\title{Simulations of a Distributed Intelligent Array Trigger for the Cherenkov Telescope Array}
\ShortTitle{Simulations of an Array Trigger for CTA}
\author{Hugh Dickinson, Frank Krennrich and \speaker{Amanda Weinstein} for the CTA Consortium\footnote{Full consortium author list at http://cta-observatory.org}\\
Iowa State University\\
Email: \email{hughd@iastate.edu},\email{krennrich@iastate.edu},\email{amandajw@iastate.edu}}
\FullConference{The 34th International Cosmic Ray Conference,\\
		30 July- 6 August, 2015\\
		The Hague, The Netherlands}

\abstract{It is anticipated that the forthcoming Cherenkov Telescope Array (CTA) will include a number of medium-sized telescopes that are constructed using a dual-mirror Schwarzschild-Couder configuration. These telescopes will sample a wide ($8^{\circ}$) field of view using a densely pixelated camera comprising over $10^{4}$ individual readout channels. A readout frequency congruent with the expected single-telescope trigger rates would result in substantial data rates. To ameliorate these data rates, a novel, hardware-level Distributed Intelligent Array Trigger (DIAT) is envisioned. A copy of the DIAT operates autonomously at each telescope and uses reduced metadata from a limited subset of nearby telescopes to veto events prior to camera readout. We present the results of Monte-Carlo simulations that evaluate the efficacy of a "Parallax width" discriminator that can be used by the DIAT to efficiently distinguish between genuine gamma-ray initiated events and unwanted background events that are initiated by hadronic cosmic rays.}

\begin{document}

\section{Introduction}\label{sec:intro}
CTA \cite{2013APh....43....3A} is a next-generation array of ground-based \gr\ telescopes that will build upon the experience of current Imaging Atmospheric Cherenkov Telescope Arrays (IACTs) including VERITAS, H.E.S.S. and MAGIC. CTA has been designed to achieve unprecedented sensitivity, angular resolution and energy resolution over a broad range of \gr\ energies. CTA will be a heterogenous array, comprising multiple telescope designs that will contribute variously to the overall performance characteristics of the array. 
A highly innovative telescope design considered for CTA uses a dual-mirror \textit{Schwarzschild Couder} (SC) configuration, which incorporates a finely pixelated camera comprising 11,328 independent silicon photomultiplier-based readout channels. The SC telescope (SCT) design provides a wide field of view ($\sim8^{\circ}$), while high-resolution capture of Cherenkov images provides excellent angular and energy resolution.

%\section{Multi-Level Triggering Strategies}\label{sec:tel_trig}

To ameliorate the effect of Night Sky Background (NSB) photons randomly triggering each of the 11,328 individual readout channels, and thereby enable a reduction of the trigger threshold, lower resolution \textit{super-pixels}  (2,832) are formed from four adjacent imaging pixels. The combined signal amplitudes of the super-pixels are used to form a Boolean-valued \textit{trigger image} with summed output threshold level corresponding to 3.1 photoelectron-equivalent (hereafter \textit{p.e.}) counts segregating true (triggered) and false values.  An \textit{individual telescope} is deemed to have generated a valid trigger condition if its Boolean trigger image includes 3 or more adjacent triggered super-pixels.
IACTs  operate in a strongly background-dominated regime. \textit{Array} triggering schemes use information from multiple telescopes to veto background events \textit{before} camera readout with the goal of stabilizing the array's energy threshold and dead time under variable ambient illumination. For the densely pixelated SCT camera, control over the array trigger rate is also desirable to guarantee that data-transfer rates remain tractable at the extremes of normal observing conditions.   

At low incident photon intensities, the overwhelming majority of single- and super-pixel triggers are associated with the ambient NSB. The rate of random super-pixel triggers is often sufficient to generate a large rate of spuriously valid individual telescope triggers. This type of background can be efficiently suppressed using a simple \textit{multiplicity array trigger} that uses temporal information to retain only those events for which multiple telescopes trigger within a nominal \textit{coincidence window}. For larger photon intensities, another background component comprising temporally correlated Cherenkov light from cosmic-ray (CR) initiated air showers becomes dominant. 

%\subsection{Computation of the Parallax Width Discriminator}\label{sec:pwidth_def}

\begin{figure}
\centering
\includegraphics[width=0.32\textwidth]{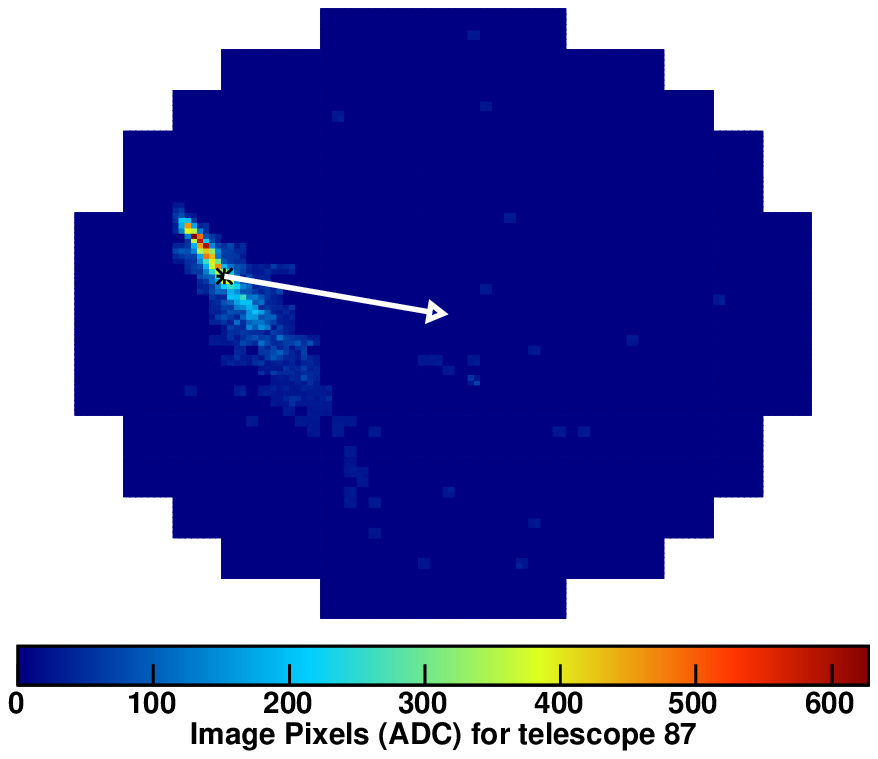}%{figures/camPlots/event_87_imPixPE}
\includegraphics[width=0.32\textwidth]{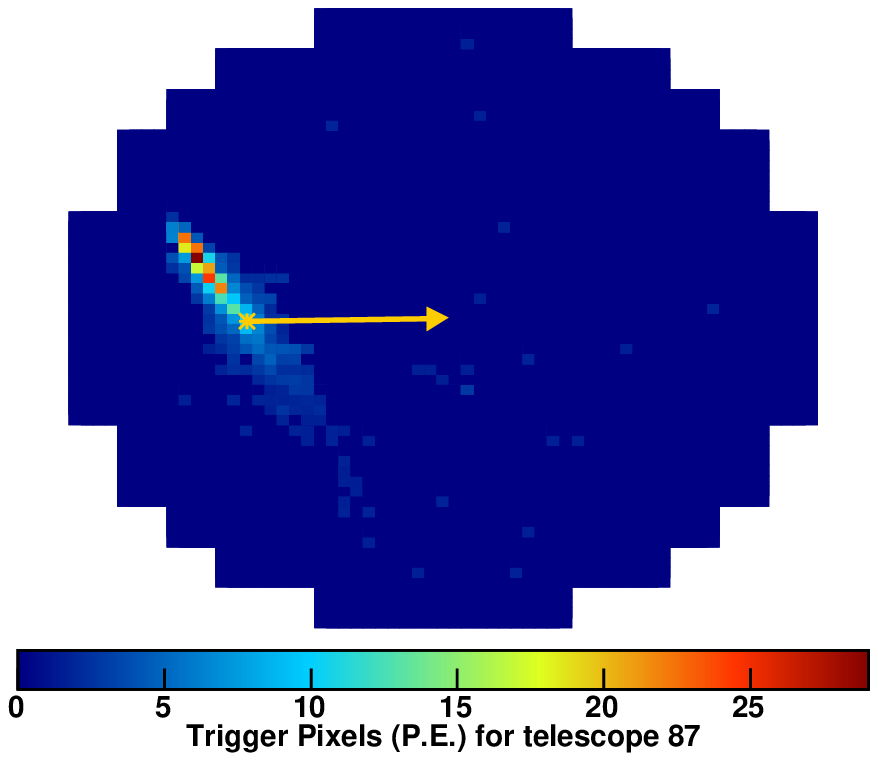}%{figures/camPlots/event_87_trigPixPE}
\includegraphics[width=0.32\textwidth]{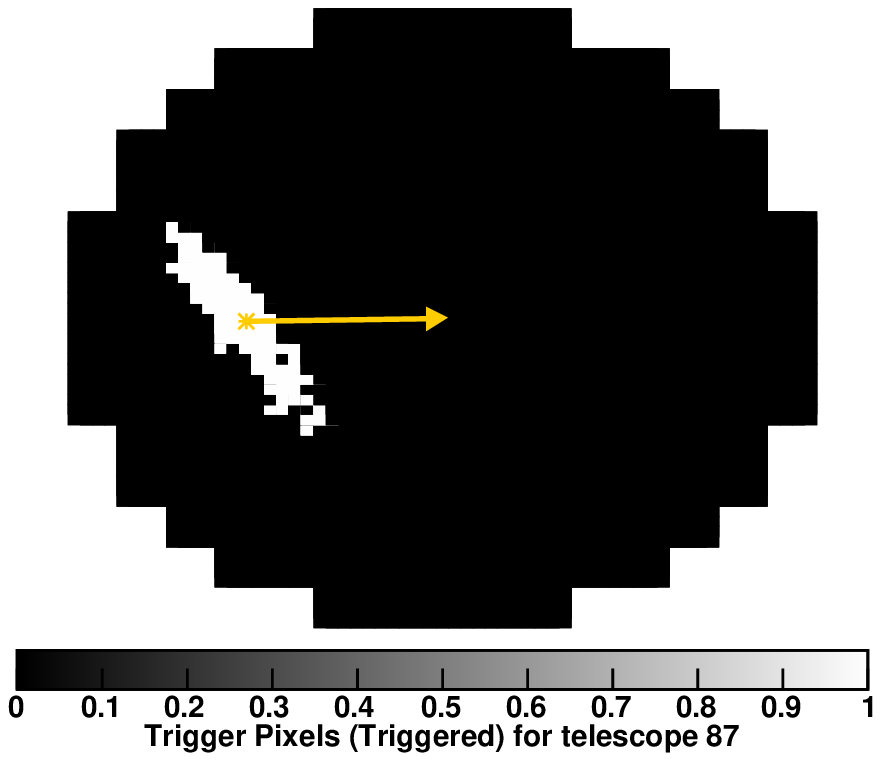}%{figures/camPlots/event_87_trigPixBool}
\caption{\small The \textit{left-hand} panel illustrates an unprocessed, finely pixelated image corresponding to a \gr-initiated air-shower.  The \textit{central} panel illustrates the intermediate, coarsely pixelated image that is derived by combining the signals from sets of 4 neighbouring imaging pixels. The \textit{right-hand} panel illustrates the Boolean-valued trigger image that is required by the algorithm that computes $P$. The \textit{orange} arrows  correspond to the vector $\mathbf{r}_{F}$, which connects the implicitly \textit{unweighted} trigger-image centroid with the \textit{fiducial} camera-plane coordinate $\mathbf{r}^{\star} = (0,0)$ at the camera centre. For comparison, the \textit{white} arrow  (\textit{left-hand} panel) connects the \textit{signal-amplitude-weighted} mean position of all \textit{imaging} pixels with $\mathbf{r}^{\star}$.}\label{fig:P_computation_Camera}
\end{figure}

Electromagnetic air-showers initiated by \grs\ are characterized by a single shower axis and triggered telescopes capture coherent elliptical images. In contrast, the hadronic component of cosmic-ray showers produces multiple sub-showers and typically yields a more fragmentary distribution of Cherenkov light. The \textit{parallax width} \cite{1995ExpAstro...6...285} discriminator ($P$) leverages the difference between CR and \gr\ shower images to rapidly distinguish between these event categories at the \textit{hardware} level.

\section{Computation of the Parallax Width Discriminator}\label{sec:pwidth_def}

It is instructive to separate the distinct algorithmic stages that comprise the computation of $P$ into two separate categories. Figure \ref{fig:P_computation_Camera} illustrates the manner in which telescope-specific camera data are processed to derive a compact geometrical representation of the captured shower image.

\begin{figure}
\center
\includegraphics[width=0.458\textwidth]{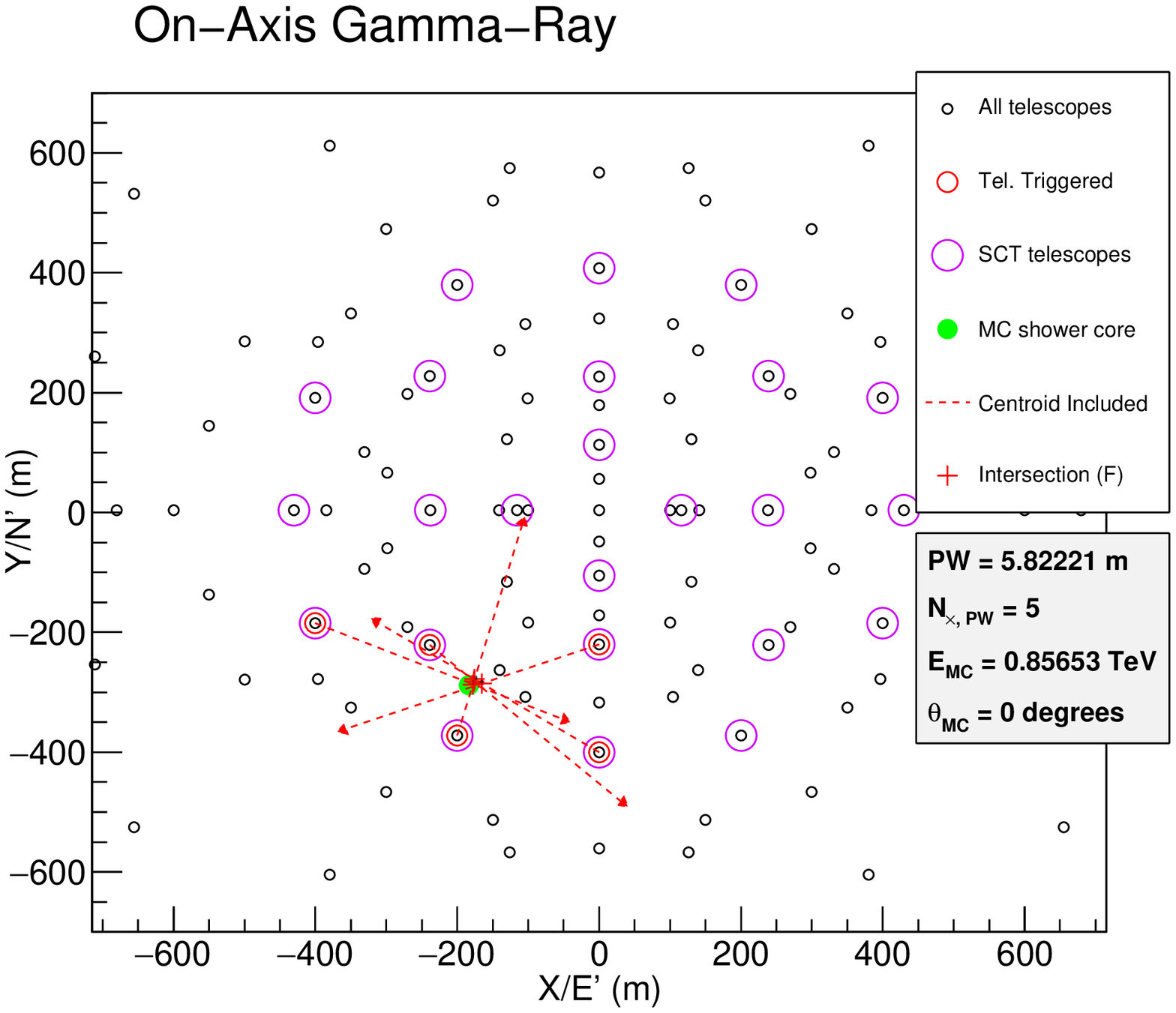}
\includegraphics[width=0.458\textwidth]{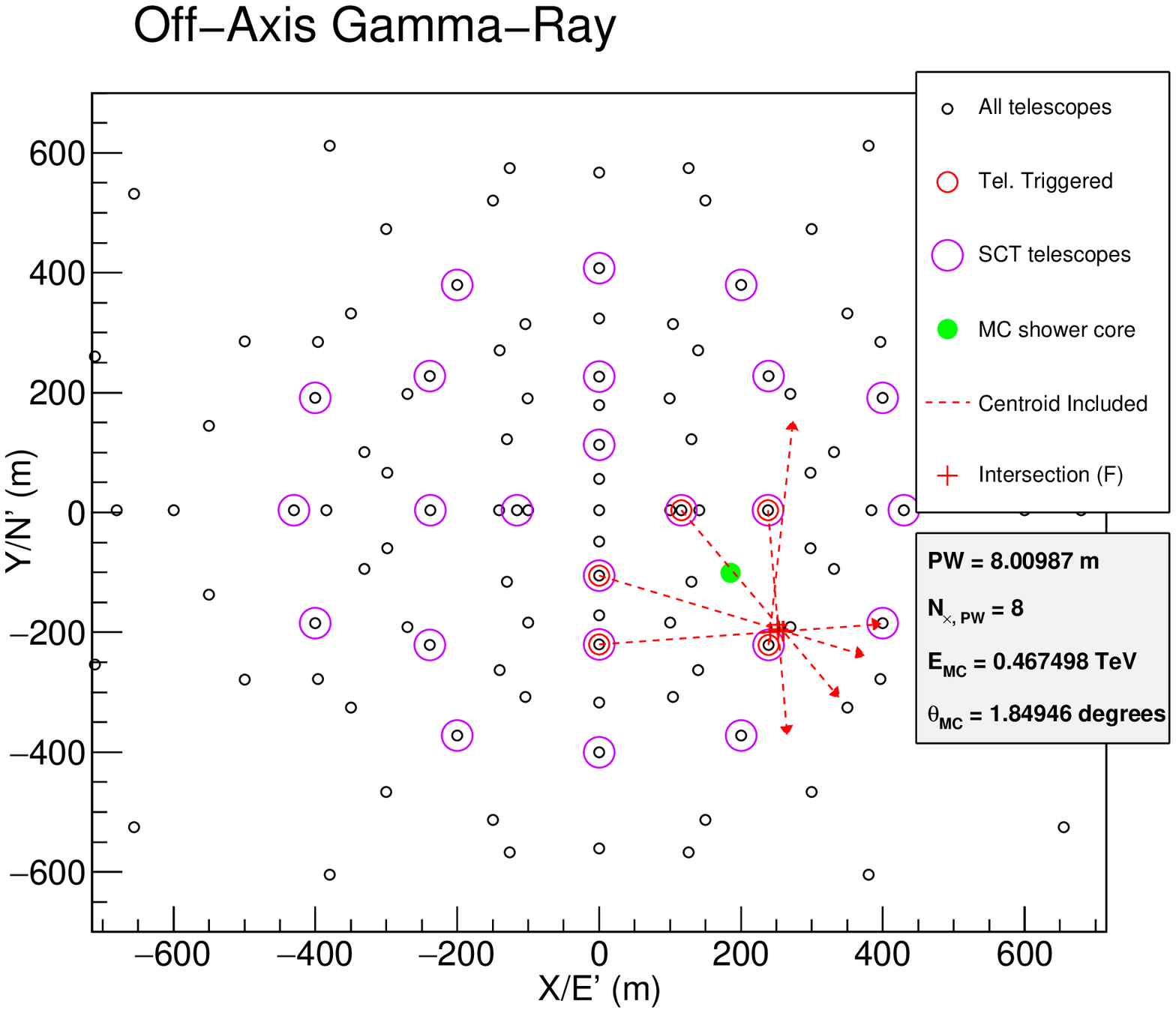}
\includegraphics[width=0.458\textwidth]{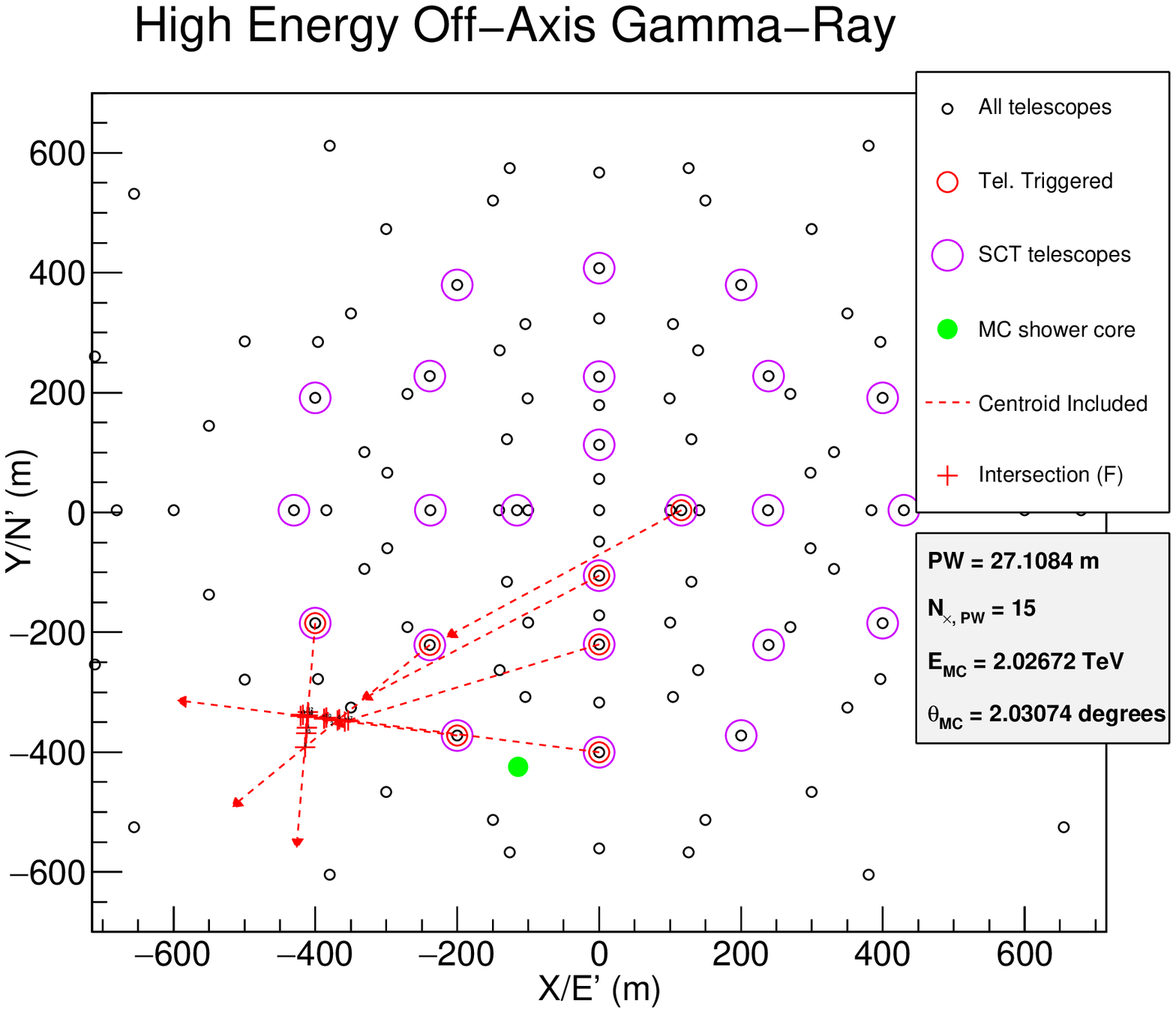}
\includegraphics[width=0.458\textwidth]{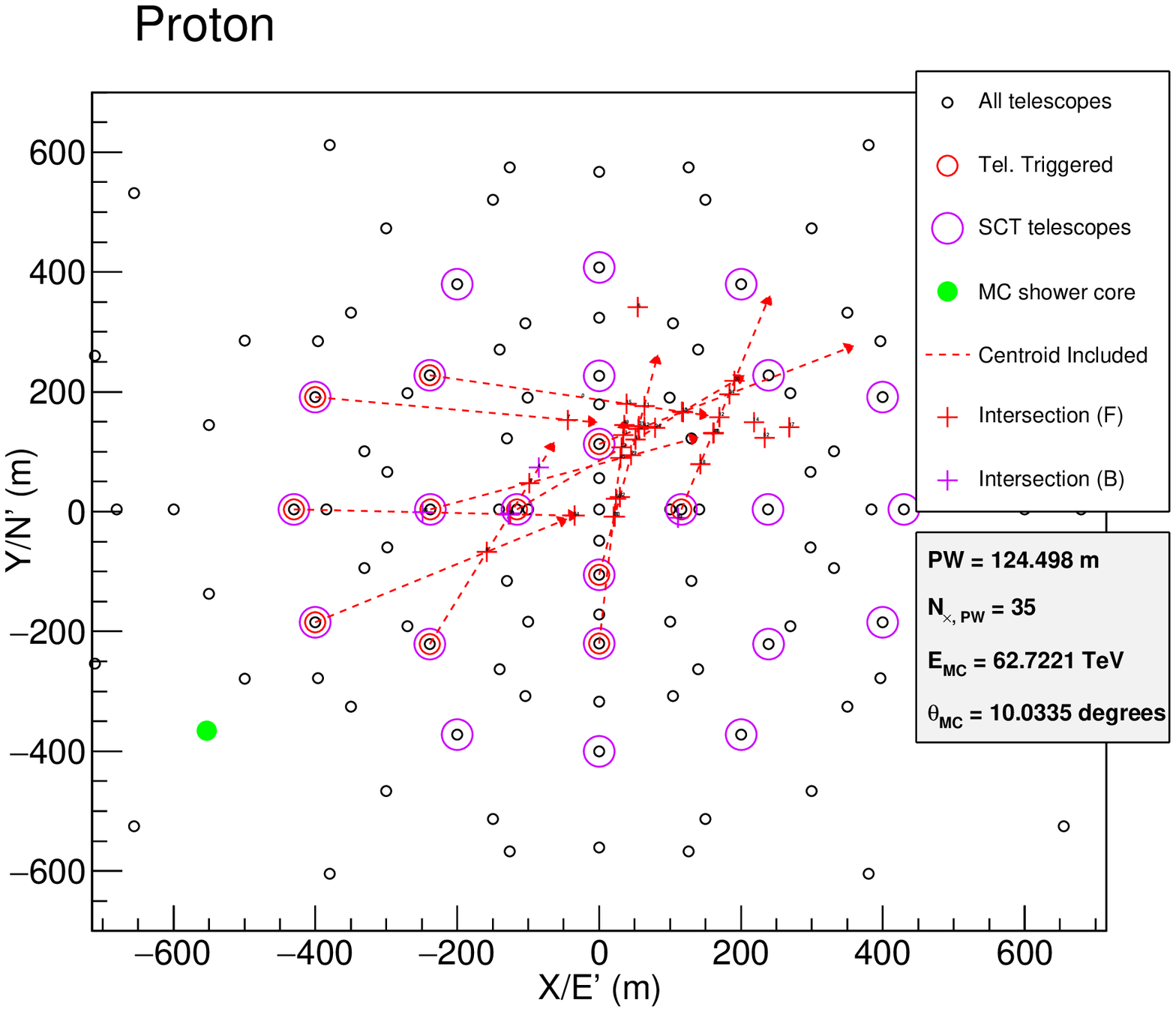}
\caption{\small Array mirror plane projections illustrating $P$ computation for an on-axis \gr\ event (\textit{top-left}), an off-axis \gr\ event with $E_{\gamma}\lesssim2$ TeV (\textit{top-right}), an off-axis \gr\ event for $E_{\gamma}\gtrsim2$ TeV (\textit{bottom-left}), and an off-axis cosmic-ray proton event (\textit{bottom-right}).}\label{fig:P_computation_MP}
\end{figure}

%\subsection{Computation of the Parallax Width Discriminator}\label{sec:pwidth_def}

\begin{enumerate}[itemsep=0ex,itemindent=0cm,leftmargin=0.5cm,topsep=0cm, parsep=0cm]
\item For each telescope that triggers, the \textit{centroid}-vector $\mathbf{r}_{C,i}$ of the coarsely-sampled trigger image is defined as the mean camera-coordinates of the set of triggered super-pixels $\mathbf{r}_{C,i} = (\langle x_{C,{\rm trig}}\rangle, \langle y_{C,{\rm trig}}\rangle)$.
\item For each centroid vector, a second vector $\mathbf{r}_{F,i} = \mathbf{r}_{C,i} - \mathbf{r}^{\star}$ is defined where $\mathbf{r}^{\star}$ is an arbitrarily selected \textit{fiducial} camera-plane coordinate $\mathbf{r}^{\star} = (x_{C}^{\star}, y_{C}^{\star})$. To simplify computation, $\mathbf{r}^{\star}$ is defined to be the camera centre $(0,0)$. 
\end{enumerate}
The events shown in each panel of Figure \ref{fig:P_computation_MP} provide representative examples of \gr\ and proton-initiated events with various simulated characteristics. They illustrate how the subsequent algorithmic operations are used to compute $P$ for each event class.   
\begin{enumerate}[itemsep=0ex,itemindent=0cm,leftmargin=0.5cm,topsep=0cm, parsep=0cm]
\item The direction of each computed $\mathbf{r}_{F}$ vector is projected from the independent coordinate systems of each telescope camera plane into a unified coordinate system spanning the \textit{mirror plane} of the telescope array, defining a set of \textit{projected} vectors $\{\mathbf{r}_{F,i}^{\prime}\}$. The mirror plane is defined to intersect the mean geographical coordinate of each telescope comprising the complete array and to lie perpendicular to their common pointing axes.
\item A set of $n_{j}$ mirror-plane coordinates $\{\mathbf{r}_{I,j}^{\prime}\}$ is computed that corresponds to the \textit{forward} intersections between the projected $\{\mathbf{r}_{F,i}^{\prime}\}$ vectors.
\item Finally, the \textit{parallax width} is defined as the dispersion of the intersection coordinates.
\begin{equation}
P = \sqrt{\frac{\sum_{j}\left| \mathbf{r}_{I,j}^{\prime} - \langle \mathbf{r}_{I,j}^{\prime} \rangle \right|^{2}}{n_{j}}}
\end{equation}
\end{enumerate}
The \textit{top-left}, \textit{top-right} and \textit{bottom-left} panels of Figure \ref{fig:P_computation_MP} all correspond to \gr-initiated events. For such events, the computed trigger-image centroids $\mathbf{r}_{C,i}$ correspond closely with the camera-plane projection of a 3-dimensional point on the air-shower axis at the height of maximal Cherenkov emission. The fiducial camera coordinate $\mathbf{r}^{\star} = (0,0)$ is also the camera-plane projection of a second three dimensional point. For \gr-initiated events, there is a single, well defined shower axis and each $\mathbf{r}_{F,i}$ connects projected 3-dimensional points that are the effectively identical for all telescopes. Accordingly, the mapping $\{\mathbf{r}_{F,i}\}\rightarrow\{\mathbf{r}_{F,i}^{\prime}\}$ yields a set of vectors which intersect at a tightly clustered region of the  array mirror plane and the computed value of $P$ is small.

Figure \ref{fig:P_computation_MP}  (\textit{top-left} ) corresponds to a \gr-initiated events aligned with the telescopes' optical axes (\textit{on-axis}). When projected into the camera plane's coordinate system, the arrival directions of on-axis events correspond precisely with the fiducial camera-plane coordinate $\mathbf{r}^{\star} = (0,0)$, which is implicitly the camera-plane projection of a 3-dimensional point \textit{on the air-shower axis}. Consequently, the derived $\{\mathbf{r}_{I,j}^{\prime}\}$ coincide closely with the projection of the axis into the array mirror plane. 
The \textit{top-right} and \textit{bottom-left} panels represent \textit{off-axis} \gr\ events with energies  below and above 2 TeV, respectively. For lower energy events, the 3-dimensional-point pairs connected by each of the $\mathbf{r}_{F}$ remain identical for all telescopes and the tight clustering of $\{\mathbf{r}_{I,j}^{\prime}\}$ is preserved. At higher energies, only one of the projected points maps to the air shower axis, so the coincidence between the  $\{\mathbf{r}_{I,j}^{\prime}\}$ and the mirror-plane projection of the air-shower axis is lost. 
%Section \ref{sec:passthrough} describes how, for $E_{\gamma}\gtrsim2$ TeV, the extensive nature of captured trigger images can disrupt the close correspondence between $\mathbf{r}_{C,i}$ and the projection of a point on the shower axis and the value of $P$ may inflate.

Finally, the \textit{bottom-right} panel illustrates a proton-initiated event. There is now no guarantee that $\mathbf{r}_{C,i}$ for each telescope corresponds with the projection of a single 3-dimensional point, since different telescopes may image multiple, different sub-showers of the hadronic cascade. Accordingly, the tight clustering of the $\{\mathbf{r}_{I,j}^{\prime}\}$ is lost and the computed value of $P$ is large.

\section{Pass-Through Trigger for High-Energy Events}\label{sec:passthrough}
The parallax width algorithm assumes that the centroids of \textit{trigger} images reliably encode the geometry of the air-showers. If the ground-plane intensity profile of Cherenkov light that is emitted by \gr-initiated air-showers  is substantially asymmetric, the validity of this assumption may degrade. Ideally, the trigger-image centroid $\mathbf{r}_{C,i}$ should correspond closely with the \textit{full image centroid} $\mathbf{r}^{\prime}_{C,i}$, defined as the signal-amplitude-weighted mean position of all \textit{imaging} pixels that trigger in response to incident Cherenkov photons. Without access to the information provided by the individual pixel \textit{amplitudes}, the Boolean-valued \textit{trigger} images appear more symmetric and the $\mathbf{r}_{C,i}$ that are used to compute $P$ may not accurately represent the air-shower geometry. 

The \textit{left} and \textit{right-hand} panels of Figure \ref{fig:P_computation_Camera} illustrate schematically how any camera-coordinate offsets $\epsilon_{C,i} = \mathbf{r}_{C,i} - \mathbf{r}^{\prime}_{C,i}$ between the two centroid definitions produce corresponding directional perturbations of each $\mathbf{r}_{F,i}$. Mapping these misaligned vectors yields a set $\{\mathbf{r}_{F,i}^{\prime}\}$ that typically increases the dispersion between the $\{\mathbf{r}_{I,j}^{\prime}\}$ intersection coordinates, and inflates the computed value of $P$. The potential magnitude of $\epsilon_{C,i}$ increases for high-energy \gr-initiated air showers, which typically produce extensive images that comprise a large number of triggered super-pixels. 

\begin{figure}
\centering
\includegraphics[width=0.438\textwidth]{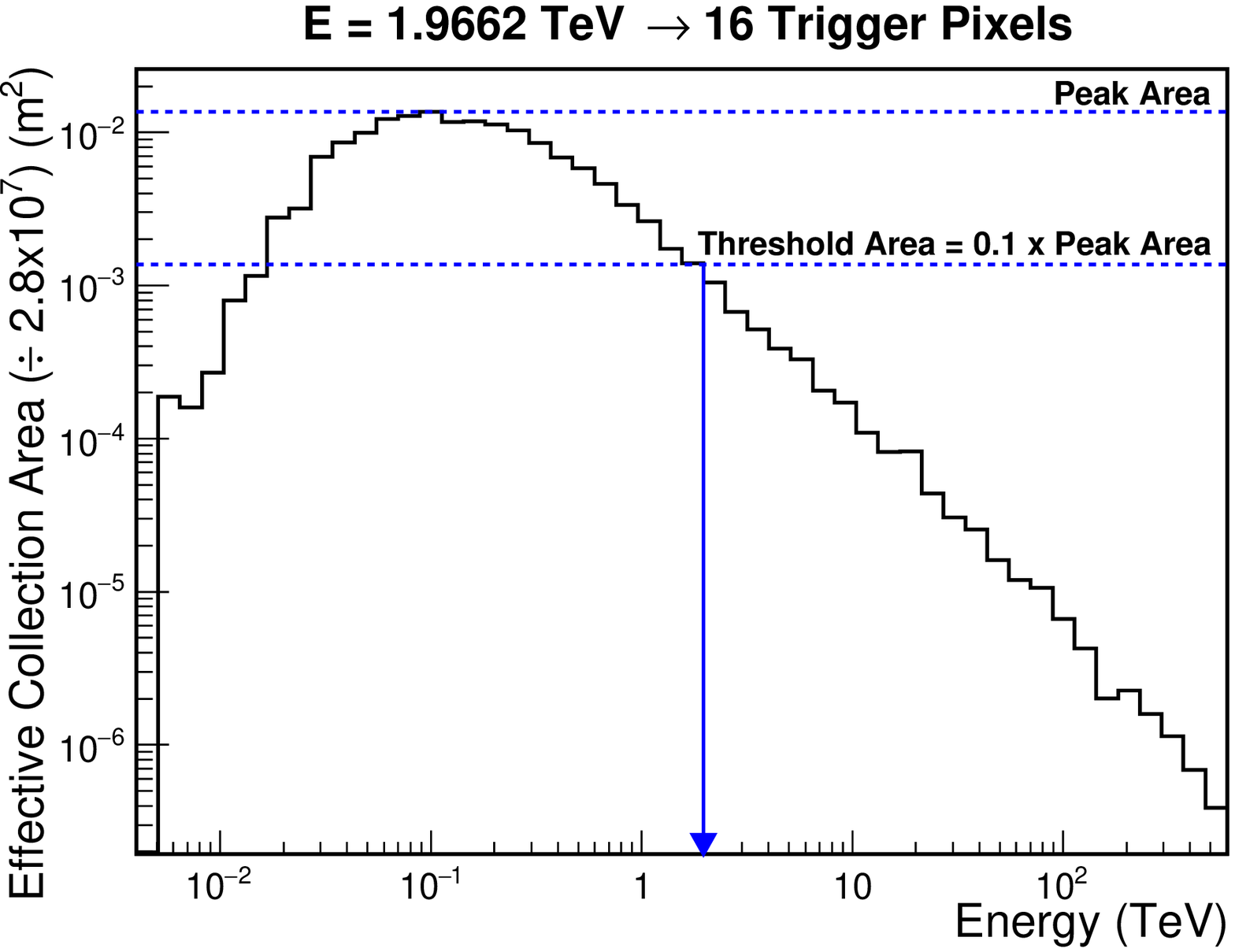}
\includegraphics[width=0.438\textwidth]{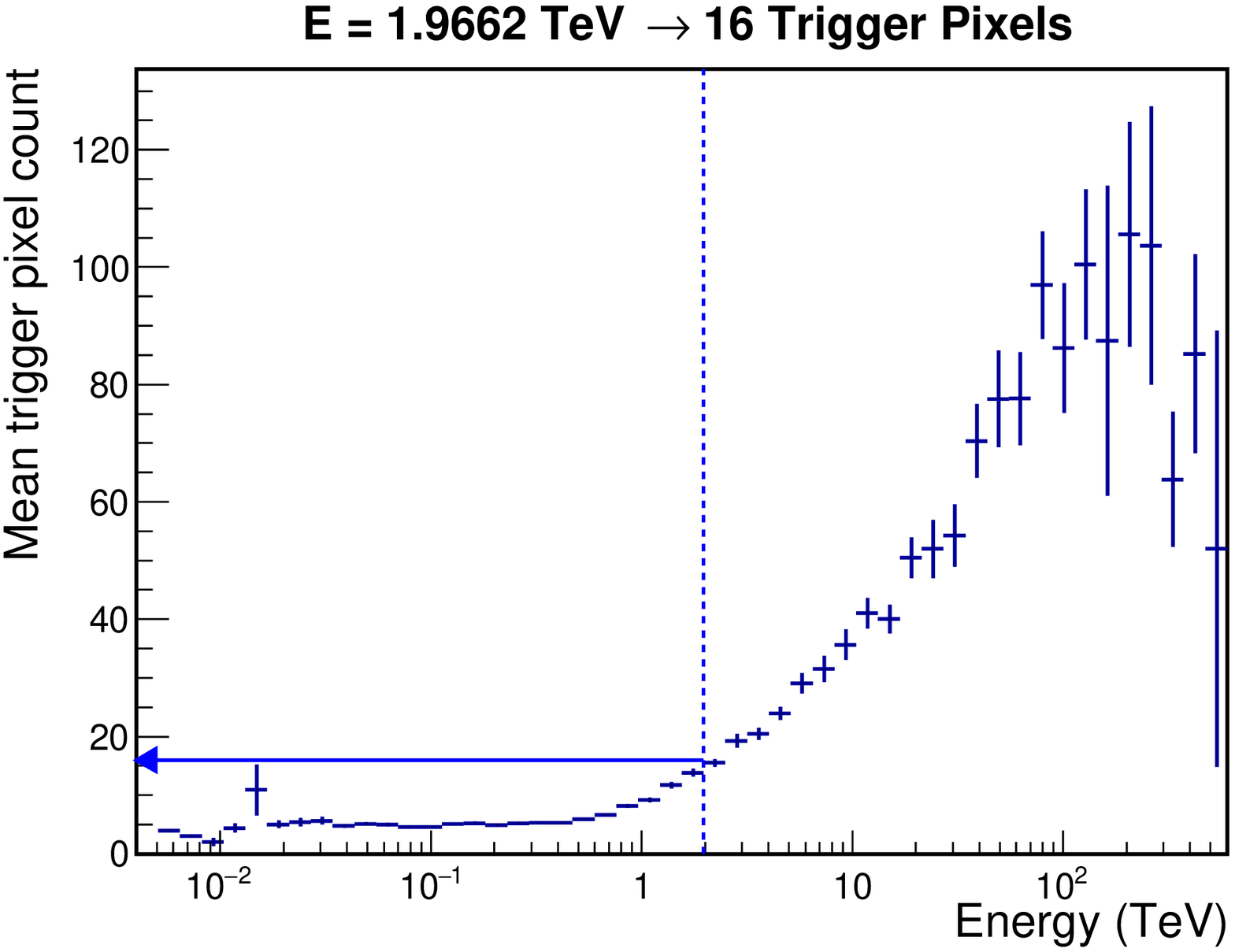}
\caption{\small Computation of the super-pixel multiplicity $n_{\rm TP}$ that is required to satisfy the pass-through criterion. In the \textit{left-hand} panel, the energy-binned effective collection area for \textit{proton}-initiated events that trigger \textit{at least one} telescope is folded with the expected cosmic ray spectral shape ($\propto E_{p}^{-2.7}$) and used to determine the typical proton energy $E_{p,10\%}$ at which the expected rate of \textit{single} telescope triggers falls below 10\% of its peak value\protect\footnotemark. In the \textit{right-hand} panel, the energy-binned distribution of trigger image super-pixel multiplicities is used to derive the value of $n_{\rm TP}$ that corresponds to events for which $E_{p}\sim E_{p,10\%}$.}\label{fig:passthrough_comp}
\end{figure}

To prevent spurious rejection of \textit{genuine} high-energy, \gr-initiated events, a pre-calibrated multiplicity threshold $n_{\rm TP}$ is used to unconditionally accept (or \textit{pass through}) events for which \textbf{any} telescope in the array captures a trigger image comprising $n_{\rm TP}$ or more super-pixels. As illustrated by Figure \ref{fig:passthrough_comp}, the expected \textit{single-telescope} trigger rate $R_{p}$ for simulated, \textit{proton}-initiated air-showers is used to calibrate an appropriate value for $n_{\rm TP}$. To retain effective suppression of the most frequent cosmic-ray triggers, a threshold corresponding to the typical super-pixel multiplicity for \textit{proton} events that trigger at 10\% of the peak single-telescope rate is adopted.

\footnotetext{Note that \gr-like images of a hadronic sub-shower typically sample a fraction of the energy of the incident proton. Accordingly, genuine \grs\ that produce images comprising $n_{\rm TP}$ trigger pixels have energies that are typically $\sim30\%$ of protons that do so.}
%\section{Simulations}\label{sec:sims}

\section{Results and Discussion}\label{sec:results}
\begin{figure}
\centering
\includegraphics[width=0.458\textwidth]{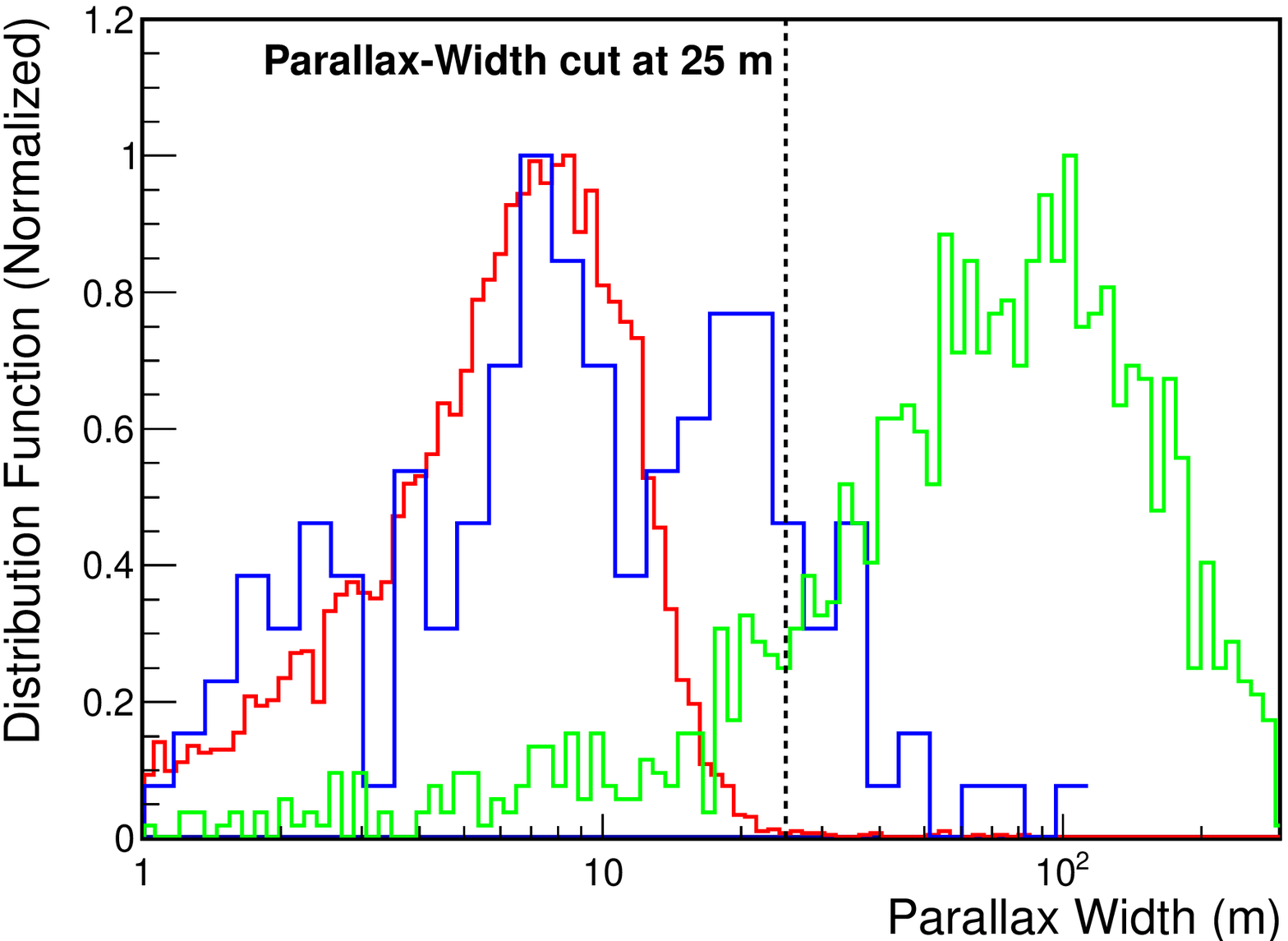}
\includegraphics[width=0.458\textwidth]{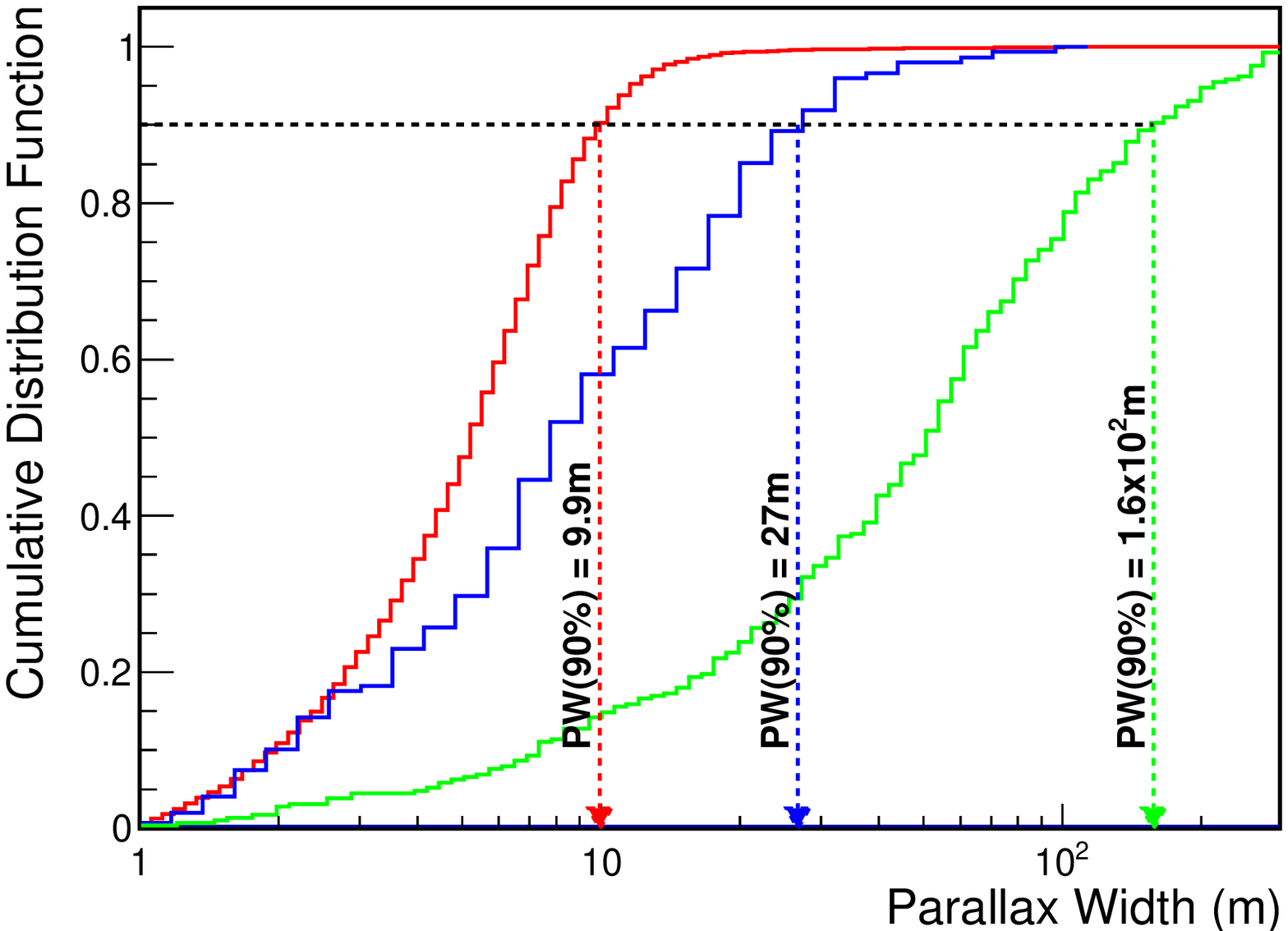}
\includegraphics[width=0.458\textwidth]{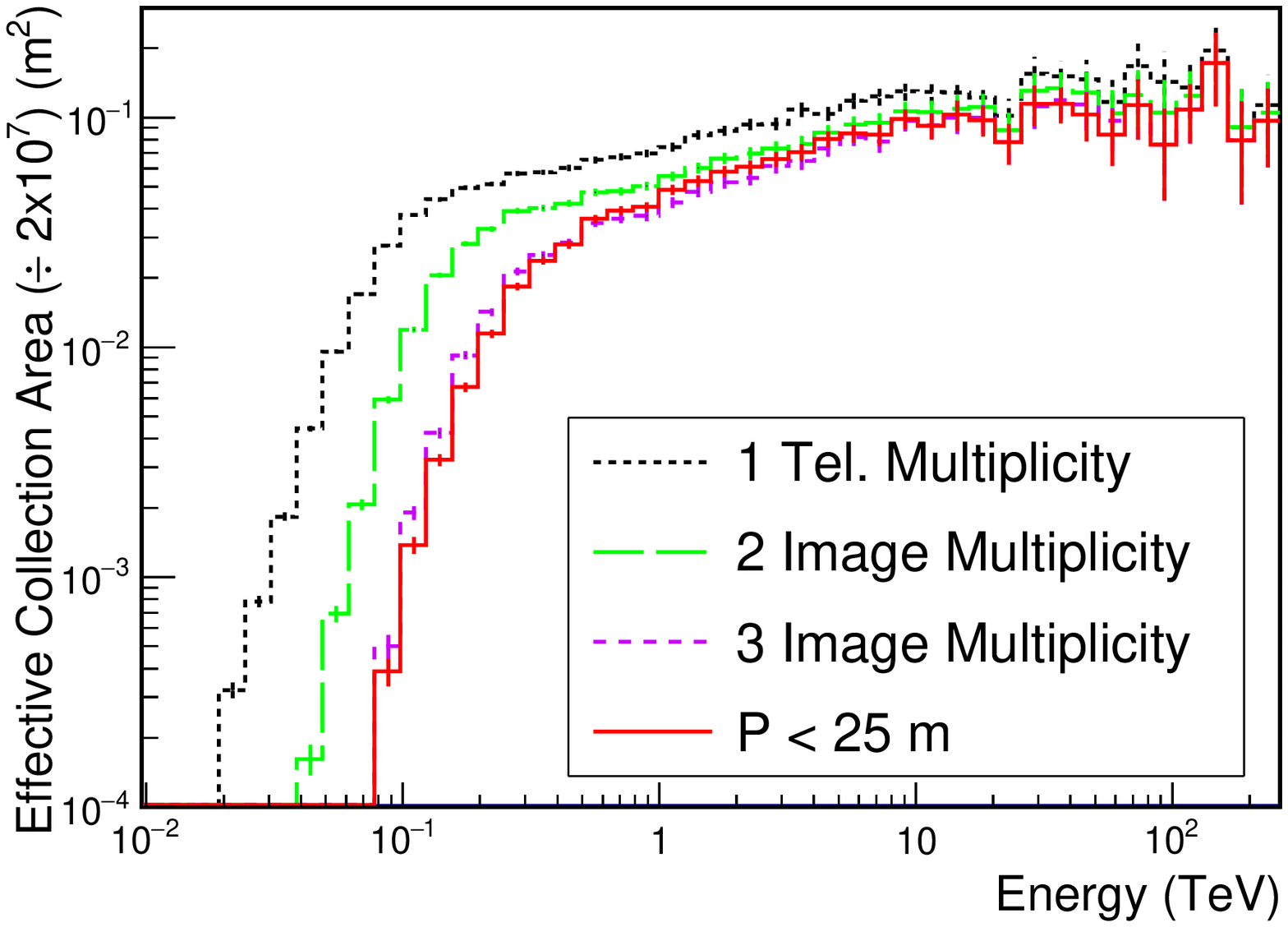}
\includegraphics[width=0.458\textwidth]{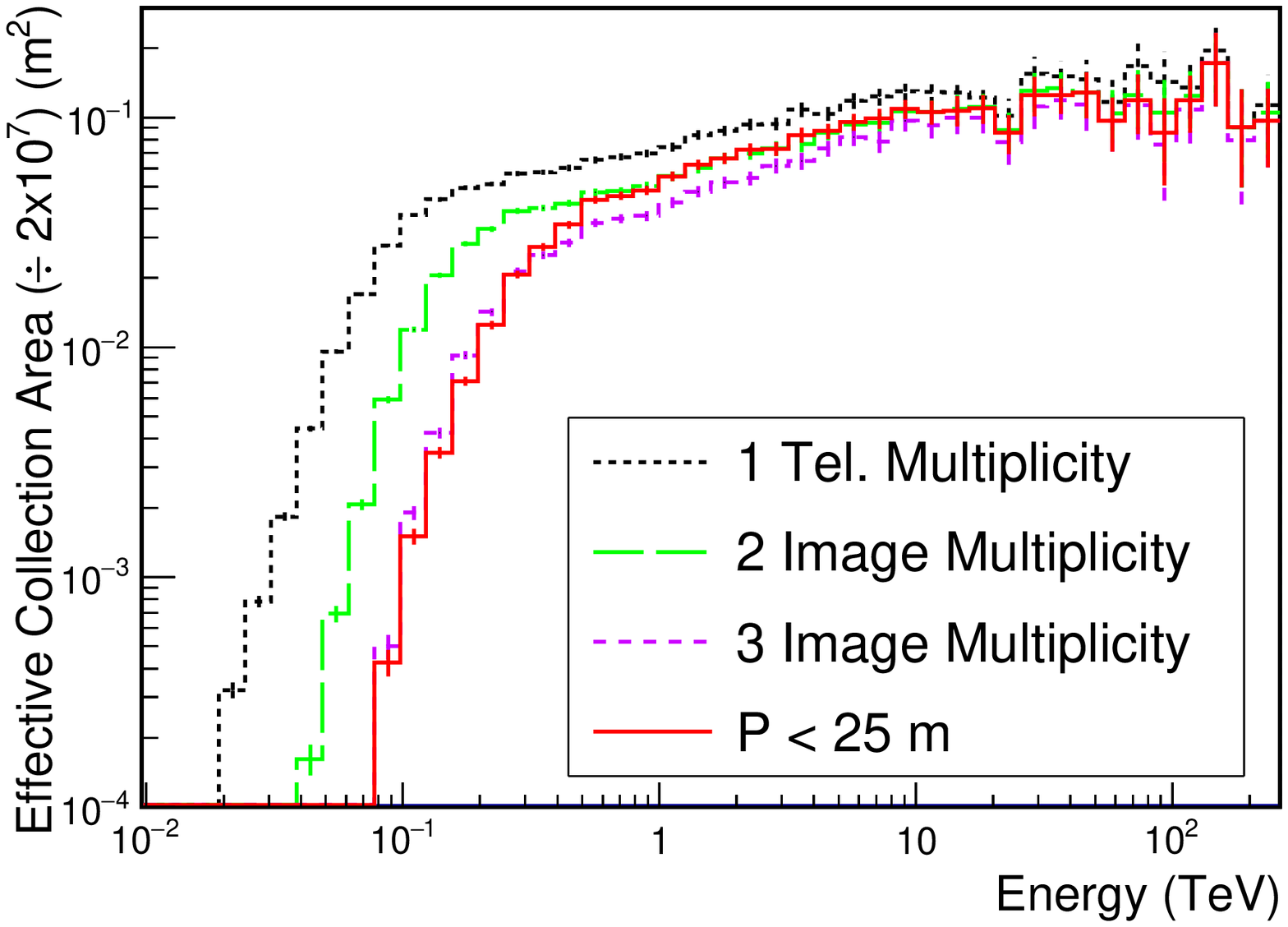}
\caption{\small \textit{Top row:} The computed distributions (\textit{left}) and CDFs (\textit{right}) of $P$ values corresponding to point-origin  (on-axis, \textit{red}) and diffuse  (\textit{blue}) \gr\ events, and diffuse cosmic-ray background events (\textit{green}). \textit{Bottom row:}  Collection areas for a point-origin \gr\ source  assuming $n_{\rm TP}\rightarrow\infty$ (\textit{left}), and  $n_{\rm TP}=16$ (\textit{right}).}\label{fig:pwidth_dists_and_gamma_aeffs}
\end{figure}

To investigate the efficacy of $P$ to distinguish \gr\- and cosmic-ray initiated air-showers, simulations of the SCT array configuration indicated by the magenta markers in Figure \ref{fig:P_computation_MP} were produced using the \texttt{sim\_telarray} software package\footnote{Simulations of point-like, extended ($2^{\circ}$ radius) \gr\ emission, and a  proton background were used. The \gr\ events model astrophysical sources located centrally in the field of view at an altitude of $70^{\circ}$ and an azimuth of $180^{\circ}$.} \cite{2008APh....30..149B}. The raw simulated camera images were provided as input to a computer simulation of the DIAT array triggering scheme outlined above.
Figure \ref{fig:pwidth_dists_and_gamma_aeffs}  (\textit{top-left}) shows the three distributions of $P$ values that are computed for each investigated dataset, while the right-hand panel displays the corresponding cumulative distributions for $P$. For the point-like \gr\ source, all the simulated photons are incident on-axis, and 90\% of computed $P$ values are  $<11$ m, in accordance with expectation. For extended \gr\ sources the majority of incident photons are incident off-axis. Nonetheless, the expected prevalence of low $P$ values is realized with 90\% of events that do \textit{not} fulfil the pass-through criterion having $P<27$ m. In contrast $\sim 75\%$ of all proton-initiated events yield $P>27$ m, confirming the expectation that $P$ is a highly efficient discriminator between \gr- and proton-initiated events.

The \textit{bottom-left-hand} panel of Figure \ref{fig:pwidth_dists_and_gamma_aeffs} plots the energy-binned effective collection areas for the simulated array configuration, assuming a point-like \gr\ source, \textit{without} application of a pass-through for events yielding extensive high-multiplicity trigger images. Curves corresponding to four distinct array-triggering criteria are shown for comparison. A fiducial scenario, corresponding to a single telescope trigger requirement (i.e. no array trigger) is shown in \textit{black}. The \textit{green} curve illustrates the effective area for events that trigger at least two telescopes and are subsequently found to yield two \textit{high-quality} Cherenkov images\footnote{\small High-quality gamma-ray images are those that are subsequently useable for parameterization and geometric reconstruction of the corresponding air shower and the properties of its progenitor \gr. Hereafter, events yielding \textit{at least} two \textit{high-quality} images will be described as \textit{minimally reconstructible}}. The \textit{violet} curve illustrates the collection area for a more desirable subset of events that yield \textit{at least three} high quality Cherenkov images. Computation of $P$ requires at least two mirror-plane intersection coordinates $\{\mathbf{r}_{I,j}^{\prime}\}$, which implicitly requires a telescope trigger multiplicity of 3 or more, but \textit{does not} require that any of the triggered telescopes yield high-quality Cherenkov images. The \textit{red} curve illustrates the collection area after application of the parallax width criterion, rejecting any events for which $P>25$ m. At low energies, the \textit{parallax-width} and \textit{three-telescope} trigger areas are comparable, but at higher energies, the parallax-width trigger retains a fraction of minimally reconstructible events that yield less than three high-quality Cherenkov images. 
%The precise criteria that define a \textit{high-quality} Cherenkov image will likely depend upon the post-observation data analysis but are typically more demanding than those required to derive a trigger-image centroid. Moreover, while the parallax-width algorithm is simple enough to be rabidly computed in hardware or firmware, real-time application of substantially more sophisticated image analyses is less feasible.
The \textit{bottom-right-hand} panel of Figure \ref{fig:pwidth_dists_and_gamma_aeffs} illustrates how adoption of a pass-through threshold  $n_{\rm TP}>16$ further enhances the retention of minimally reconstructible \gr\ events by the DIAT.

\begin{figure}
\centering
\includegraphics[width=0.438\textwidth]{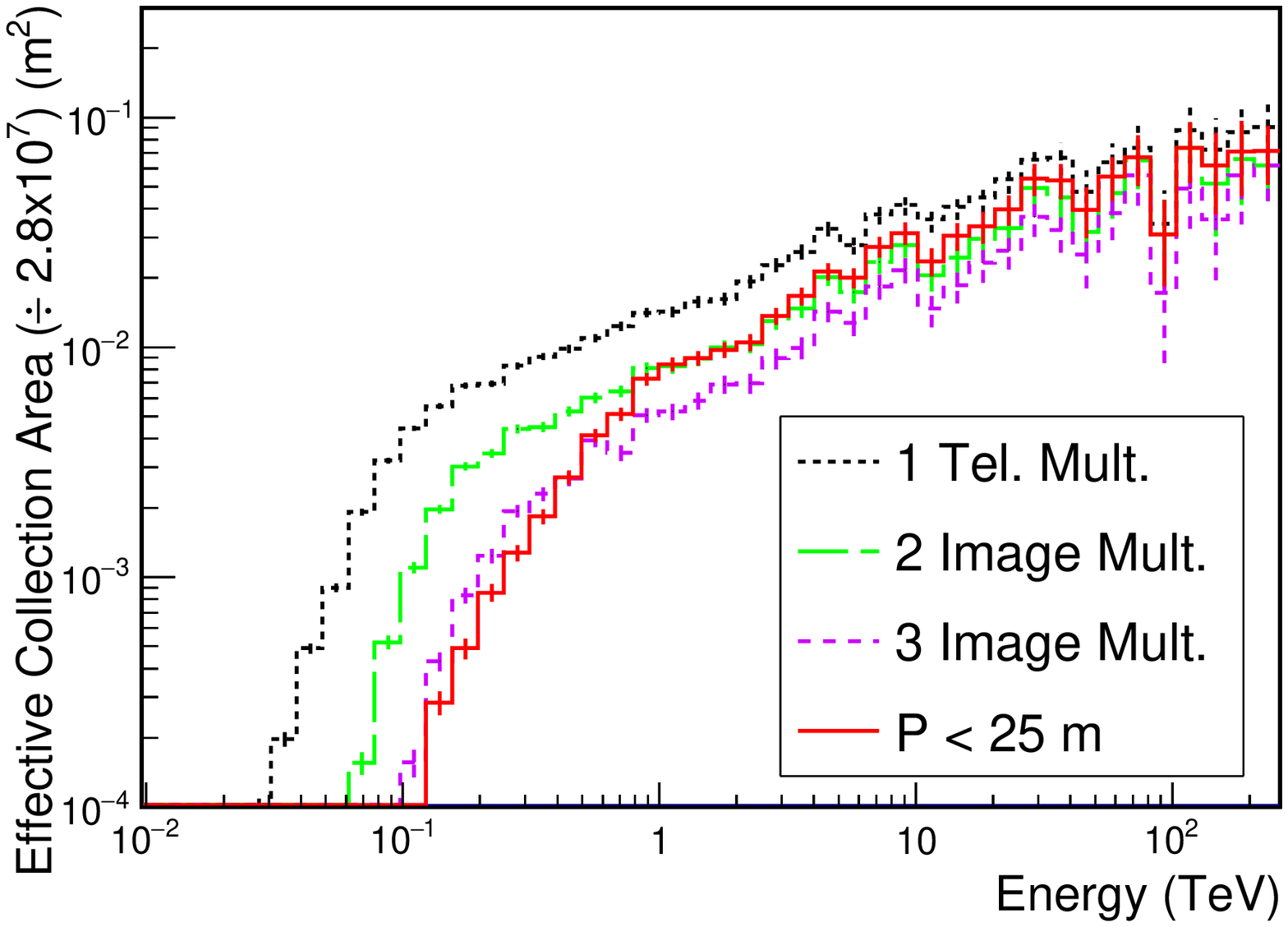}
\includegraphics[width=0.438\textwidth]{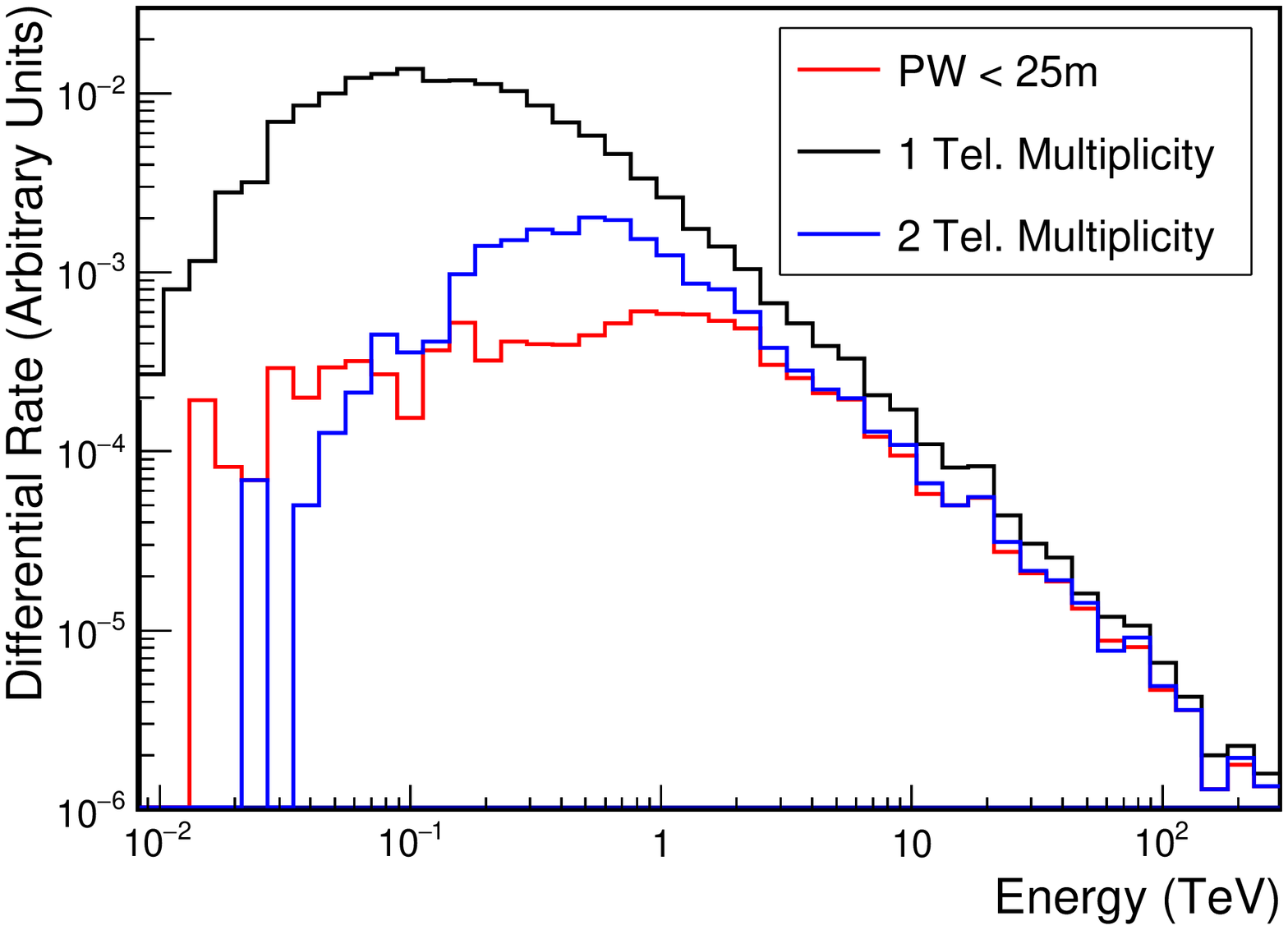}
\includegraphics[width=0.438\textwidth]{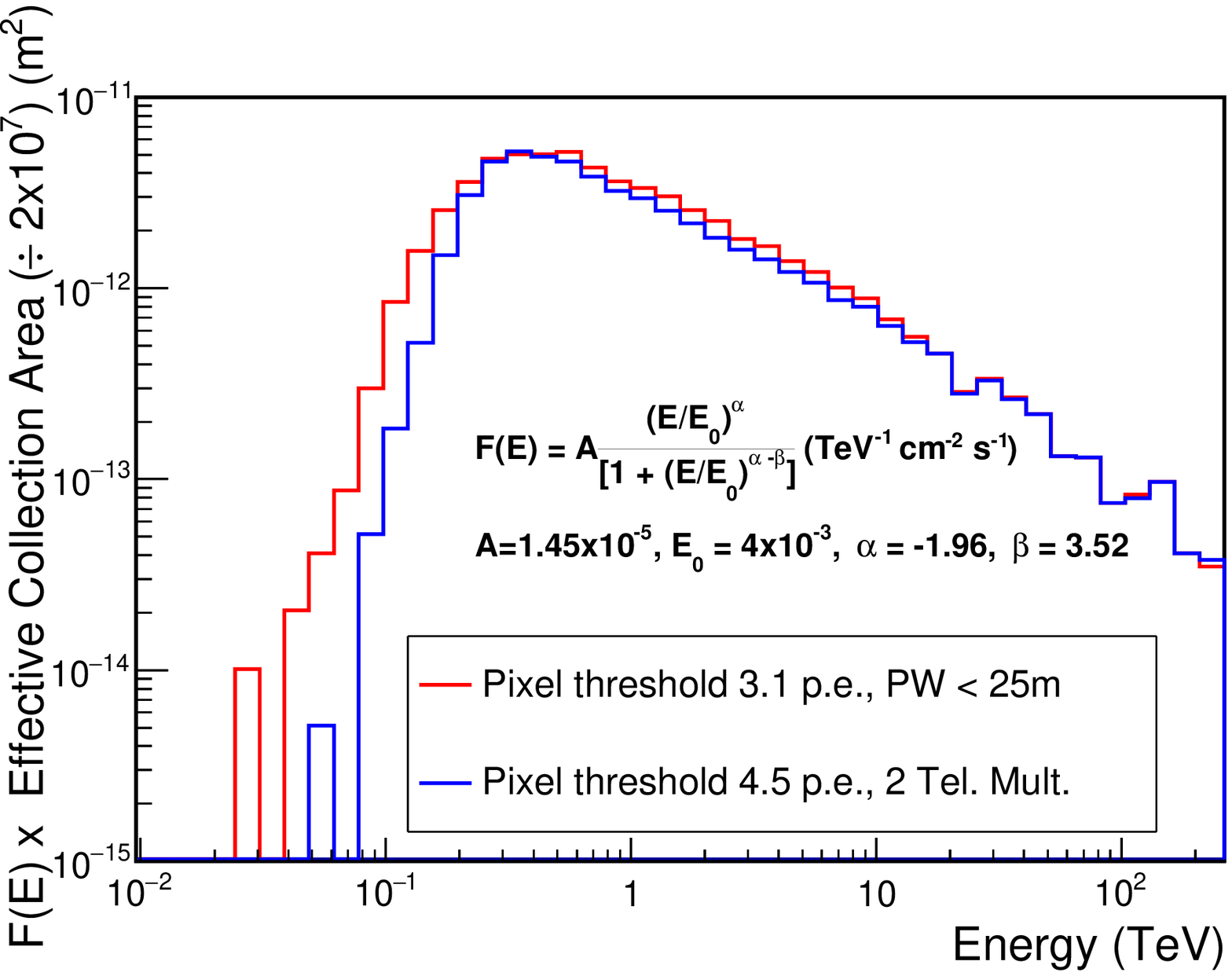}
\includegraphics[width=0.438\textwidth]{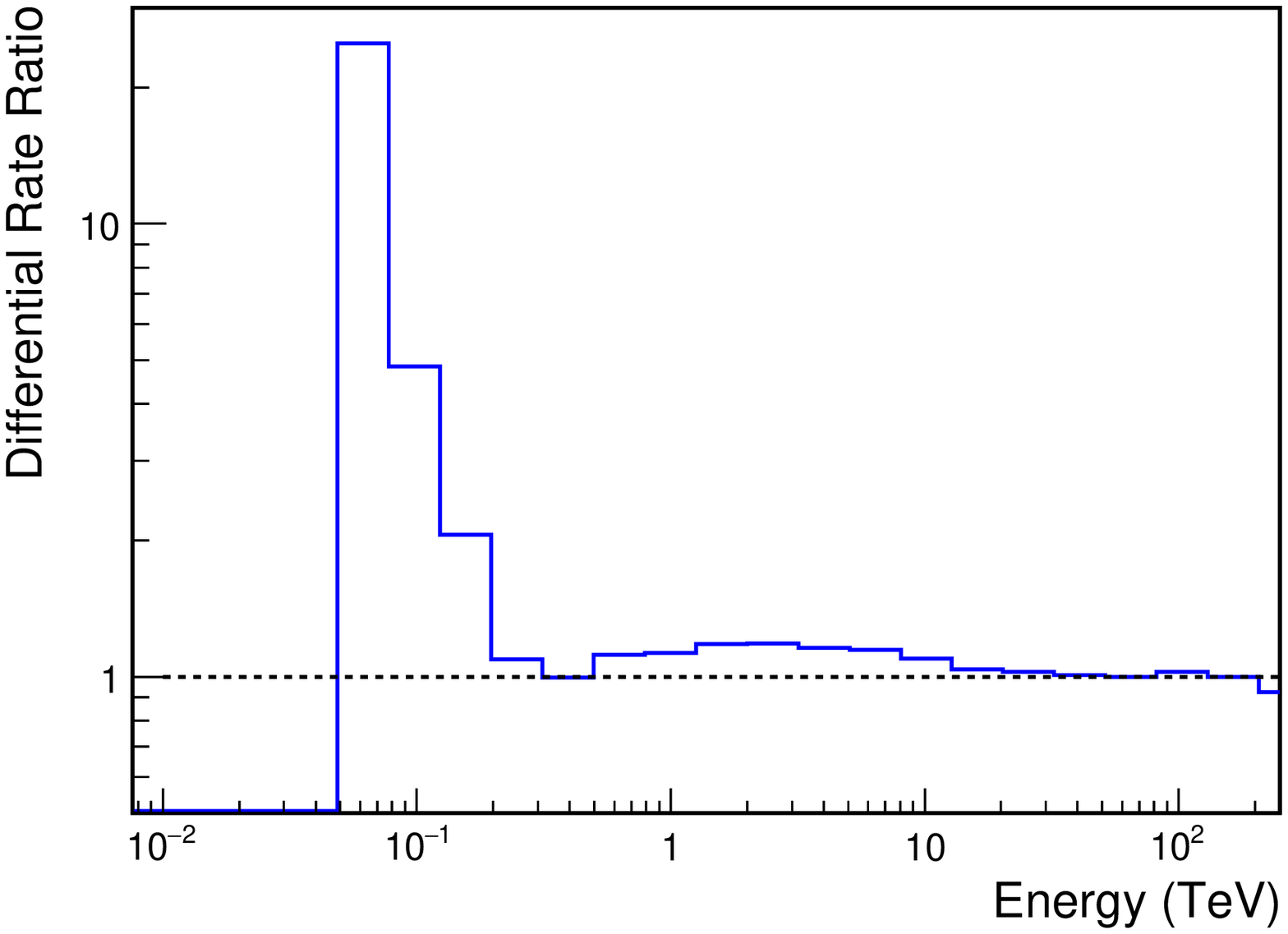}
\caption{\small \textit{Top row:} Collection areas for diffuse \gr\ emission (\textit{left}), and diffuse protons (\textit{right}) corresponding to a pass-through  super-pixel multiplicity threshold $n_{\rm TP}=16$. \textit{Bottom row:} The expected array trigger rates for on-axis \gr-initiated events (\textit{left}) are shown for  a Crab-pulsar-like spectral shape. The red curve corresponds to application of the parallax width algorithm with $P<25$ m, a pass-through threshold $n_{\rm TP}>16$, and a pixel trigger threshold corresponding to 3.1 p.e. per trigger pixel. For comparison, the \textit{blue} curve plots the expected array rate for 
a traditional two-telescope multiplicity array trigger, with a  trigger-pixel threshold of 4.5 p.e.
%\textit{minimally reconstructible} on-axis \gr-initiated events (i.e. events that yield at least two useable Cherenkov images) for a pixel trigger threshold corresponding to 4.5 photoelectrons per trigger pixel. 
Furthermore, the ratio of the \textit{red} and \textit{blue} curves (\textit{right}) are shown, indicating that application of the parallax width discriminator enables effective recovery of low energy events.}\label{fig:event_recovery_gamma-diff_and_proton_aeffs}
\end{figure}

The \textit{top-left-hand} panel of Figure \ref{fig:event_recovery_gamma-diff_and_proton_aeffs} uses simulated effective collection areas to demonstrate the response of the DIAT to an extended astrophysical \gr\ source, adopting $n_{\rm TP}>16$. At low energies the rejection of events yielding $P>25$ m retains a large fraction of events that yield three high quality Cherenkov images, while at energies $\gtrsim1$ TeV, the pass-through trigger results in retention of \textit{all} minimally reconstructible events. The \textit{top-right-hand} panel of Figure \ref{fig:event_recovery_gamma-diff_and_proton_aeffs} illustrates the power of $P$ to effectively reject cosmic-ray proton-initiated events. The \textit{blue} curve illustrates the effective collection area for a traditional two-telescope multiplicity array trigger, that accepts events for which at least two neighbouring telescopes trigger. Both array triggering strategies reduce the energy-integrated collection area by a factor $\sim 7$, but it is evident that the parallax-width trigger outperforms the two-telescope multiplicity requirement at lower energies where the incident cosmic ray rate is largest.
Figure \ref{fig:event_recovery_gamma-diff_and_proton_aeffs} (\textit{bottom row}) illustrates how DIAT using the parallax width algorithm can be used to improve the low-energy sensitivity of the SCT subarray. Field-to-field variability of the NSB intensity between observations can induce unpredictable spikes in the array trigger rate if accidental coincidences between spurious single telescope triggers cannot be effectively suppressed. Typically, the required suppression is achieved by requiring a higher single-pixel trigger threshold, which increases the overall low-energy threshold of the array. The ability of the DIAT to perform real-time multi-telescope event vetoing allows lower pixel thresholds to be used while maintaining tractable array trigger rates. In Figure \ref{fig:event_recovery_gamma-diff_and_proton_aeffs}, the differential trigger rates for a Crab-pulsar-like \gr\ source 
%\cite{2011Sci...334...69V} 
are compared for events that satisfy $P<25$ m assuming the nominal trigger pixel threshold of 3.1 p.e., with the set of events that 
fulfil a traditional two-telescope multiplicity array trigger.
%are \textit{minimally reconstructible} assuming a higher 4.5 p.e. pixel trigger threshold. 
The lower pixel threshold made feasible by a hardware-level DIAT yields a substantial improvement in sensitivity below 200 GeV, achieving a factor of $\sim20$ enhancement at the lowest energies.

% \section{Conclusions}\label{sec:conclusions}

In summary,  simulations show that efficient rejection of cosmic-ray events is possible with a hardware array trigger using the moments of Cherenkov light images. The algorithm successfully discriminates between background and genuine \gr\ triggers, retaining a large majority of  reconstructible events. By vetoing spurious single and multiple telescope triggers before data are read out, the algorithm reduces the array trigger rate, enabling finely sampled events with heavy data payloads to be generated by participating SCTs without overwhelming the array data transfer infrastructure. Real-time consideration of data from multiple telescopes also allows the rate of NSB-induced array triggers to be controlled without increasing the single-pixel trigger thresholds. The resultant enhancement of low-energy sensitivity may be particularly useful for studies of spectrally soft targets like the Crab pulsar.
\vspace{-5pt}\paragraph{Acknowledgements:} We gratefully acknowledge support from the agencies and organizations 
listed under Funding Agencies at this website: http://www.cta-observatory.org/.
\vspace{-10pt}\

%\bibliographystyle{ICRC2015/JHEP}
%\bibliography{ParallaxWidthICRC}
\end{document}